\documentstyle[12pt]{article}
\newcommand{\beq}{\begin{equation}}
\newcommand{\eeq}{\end{equation}}
\newcommand{\beqy}{\begin{eqnarray}}
\newcommand{\eeqy}{\end{eqnarray}}
\newcommand{\beqyy}{\begin{eqnarray*}}
\newcommand{\eeqyy}{\end{eqnarray*}}

\def\p{\partial}

\def\ap{\alpha'}
\catcode`\@=11
\def\numberbysection{\@addtoreset{equation}{section}
\def\theequation{\arabic{section}.\arabic{equation}}}
\def\thesection{\arabic{section}.}

\def\appendix{\setcounter{section}{0}
        \def\thesection{Appendix \Alph{section}}
        \def\theequation{\Alph{section}.\arabic{equation}}}
\begin{document}
\pagestyle{empty}
\begin{flushleft}
{\it  Yukawa Institute Kyoto}
\end{flushleft}
\begin{flushright}
 YITP-99-26    \\
{\tt  hep-th/9905055}\\
 May 1999      \\
\end{flushright}
\renewcommand{\thefootnote}{\fnsymbol{footnote}}
\vfill
\begin{center}
{\LARGE{\bf 
D0-branes in an H-field Background and \\ 
\vspace{2mm}
Noncommutative Geometry \\
} }
\vspace{15mm}
Masahiro Anazawa 
\\
\vspace{3mm}
      {\it  Yukawa Institute for Theoretical Physics \\
        Kyoto University, Kyoto 606-8502, Japan\\  } 
\vspace{3mm}
{\tt anazawa@yukawa.kyoto-u.ac.jp}
\end{center}
\renewcommand{\thefootnote}{\arabic{footnote}}
\setcounter{footnote}{0}
\vspace{10mm}
\begin{abstract}
\baselineskip 17pt
It is known that if we compactify D0-branes on a torus with
constant B-field, the resulting theory becomes SYM  theory 
on a noncommutative dual torus. We discuss the extension to the case
of an H-field background. In the case of a constant H-field on 
a three-torus, we derive 
the constraints to realize this compactification
by considering the correspondence to string theory. We carry out 
this work as a first step to examine the possibility to describe 
transverse M5-branes in Matrix theory.
\end{abstract}
\vfill
\begin{flushleft}
\end{flushleft}
\newpage
\pagestyle{plain}
\setcounter{page}{1}
\numberbysection
\baselineskip 17pt
\renewcommand{\arraystretch}{1.5}
\section{Introduction}
\label{sec:intro}
\indent

Since the proposal of Matrix theory \cite{BFSS}
as a nonperturbative formulation of  M-theory in the infinite momentum 
frame or DLCQ of M-theory \cite{Susskind}, 
it has passed various kinds of  consistency tests \cite{Review}.
Because the action of Matrix theory is equivalent to a regularized 
action of supermembrane in the light cone gauge \cite{DWHN},
the relation of membranes to Matrix theory is clear, and
fundamental strings can also be described in Matrix theory \cite{string}.

As for M5-branes, the situation is more obscure \cite{BD,BSS,GRT}.
In \cite{BSS}, the supersymmetry algebra was examined for finite $N$
Matrix theory, and various brane charge densities were identified.
While the charge density of longitudinal M5-brane wrapping 11th 
direction really emerged as a central charge, that of transverse 
M5-brane wasn't found.
In \cite{EMM}, a matrix model describing open membranes was formulated
in the light cone gauge, 
motivated by the idea to describe M5-branes as boundaries on which open
membranes can attach. However, there was also difficulty in describing 
transverse M5-branes. 
These situations are very unsatisfactory form the viewpoint of 
11-dimensional Lorentz symmetry. In general, brane charges are expected
to merge in the large $N$ limit, where 11-dimensional Lorentz symmetry
should be hold. Then we consider that there still remain  possibilities
to describe transverse M5-branes in Matrix theory, if we consider
some particular large $N$ limit.

On the other hand, it was shown that the usual compactification of
 Matrix theory on tori \cite{Taylor,GRT} can be extended to 
 noncommutative cases \cite{CDS}.
  In the usual compactification on $T^d$, the resulting 
 theory becomes SYM theory on a dual torus. This can be extended to 
 SYM theory on a noncommutative dual torus. There, 
 the parameters of the noncommutativity were constant,  and their 
 physical interpretation was argued to be  the fluxes 
 of the three-index gauge field $C_{-IJ}$ \cite{CDS}.
 This field corresponds to $B_{IJ}$ in type IIA string theory.
 This noncommutative nature was derived in \cite{DH} -\cite{Schomerus}
 from various points of view.
 
 Since this fact shows that we can deal with $C_{-IJ}$ background in 
 Matrix theory, this leads us to an another possibility of a description
 of transverse M5-branes.
 If we succeed to find a way to describe the backgrounds corresponding
 to non-zero transverse M5-brane charge in Matrix theory,
 the resulting theory might describe transverse M5-branes.
 Since transverse M5-branes correspond to NS5-branes in type IIA string 
 theory, for example, we need to consider a background satisfying
 \beq
  \frac{1}{2\pi\ap}\oint_{S^3} H= 2\pi n \; .
 \label{eq:charge}\eeq
 To do this, we have to find a way   to treat space dependent 
 $B_{IJ}$.\footnote{In \cite{Compean}, an extension  to
  non-constant $B_{IJ}$ was discussed using deformation
 quantization theory. There, $B_{IJ}$ depends on the position of the
 dual compact space. However, we need $B_{IJ}$ depending on the position
 of the original compact space $T^3$ not but dual space.}
 Instead of eq.(\ref{eq:charge}), as a first step in this direction,
  we will consider a background satisfying
 \beq
  \frac{1}{2\pi\ap}\oint_{T^3} H = 2\pi n \;,
 \label{eq:charge2}\eeq
 because this is the simplest and nontrivial extension of the 
 case in \cite{CDS}.
 
 In this paper, we will consider a low energy system of D0-branes, 
 and discuss how to realize the compactification on
  $T^3$ with
 the background $B_{IJ}$ satisfying eq.(\ref{eq:charge2}).
 We assume that this compactification can be realized in Matrix theory with
 suitable constraints on the matrices similar to those in \cite{Taylor}.
 We will discuss what constraints we need from the correspondence
 to string theory in a similar way to \cite{CK}.
 We will find that the center of mass coordinates of D0-branes no longer 
 decouple from the remaining degrees of freedom.
 We also discuss the relation to the ordinary noncommutative torus
 compactification.

\section{Three-torus with an H-field background}
\label{sec:background}
\indent

Let us consider the compactification of  Matrix theory
on a rectangular torus.
For flat background, the Lagrangian of Matrix theory is given by the 
low energy Lagrangian of D0-branes in ten dimensions,
\beqy
&&{\cal L}=\frac{1}{2 g_s \sqrt{\ap}} {\rm Tr}
 \left\{  (D_t X^{I})^2
 + \frac{1}{2(2\pi \ap)^2}[X^{I},X^{J}]^2 
 \right. \nonumber \\
  &&\qquad \qquad \left.
 +\frac{i}{2\pi\ap}\Theta^T  D_t \Theta-\frac{1}{(2\pi\ap)^2} \Theta^T
 \Gamma_{I}[X^{I},\Theta] \right\} , 
\label{eq:MT_action}
\eeqy
where $I,J=1,\cdots,9$.
Here, the appropriate $\ap \to 0$ limit  specified in \cite{Seiberg,Sen}
 must be taken
to make this system  correspond to M-theory in 11 dimensions.
As it is well known, for the usual compactification on $T^d$ with 
radii $R_i$, we should impose the quotient condition on the matrices
 \cite{Taylor,GRT},

\beq
\begin{array}{rclc}
U^{-1}_i X^{I}U_i & = & X^{I}+2\pi R_i \delta_{I,i} & \quad i=1,\cdots,d \\
U^{-1}_i \Theta U_i & = & \Theta   \;, & 
\end{array}
\label{eq:usual_QC}\eeq
where $U_i$ are unitary operators.
This relation means that $X^{i}+2\pi R_i$ is equivalent to $X^i$ up to 
a unitary transformation. From the consistency of these relations, 
 $U_iU_jU^{-1}_iU^{-1}_j$ must commute to $X^{I}$ and $\Theta$, 
 so we obtain \cite{CDS}
\beq
U_i U_j=e^{2\pi i \theta_{ij}} U_j U_i \;.
\label{eq:NC_algebra}
\eeq
For the case where $\theta_{ij}$ are constant, their physical interpretation
is argued to be the fluxes of $B_{ij}$ integrated on  the $T^2$ extending
 $x^i$ and $x^j$ directions \cite{CDS}-\cite{Schomerus}. 

In this paper we would like to consider a $T^3$ compactification with 
a non-zero $H_{123}$ background, which is topologically quantized as
\beq
T_2 \oint_{T^3} H = 2\pi n \;,
\eeq
where $T_2$ is  $1/2\pi\ap$  and $n$ is an  integer.
In this case,  $B_{ij}$ are not constant and depend on the position
 in $T^3$, so the above 
procedure of toroidal compactification can not be applied.
The quotient condition must be modified. In this paper we will discuss 
how to modify the usual quotient condition to describe the torus 
compactification with non-zero $H$ field. 
Throughout the paper, we mainly consider the case of a D0-brane
compactified on $T^3$, because the extension to N D0-branes is 
straightforward.

Let us assume that $(x^1,x^2,x^3)$ directions are compactified on $T^3$ 
and we have  constant $H_{123}$. For simplicity we take 
 $B_{ij}$ as
\beq
B_{12}(x^3)=\frac{1}{(2\pi)^3 R_1 R_2 R_3} \frac{2\pi n}{T_2} x^3 \;,
\eeq
and $B_{23}=B_{31}=0$.
This configuration is topologically nontrivial and the boundary condition
of $B_{ij}$ is specified introducing a nontrivial gauge transformation
\beq
B(x^3+2\pi R_3)=B(x^3) +d \Lambda^{(1)} \;,
\label{eq:BCforB}\eeq
where $\Lambda^{(1)}$ is a 1-form field. To be specific, we take as
\beq
{\Lambda^{(1)}_1 \choose \Lambda^{(1)}_2}
={0 \choose
\frac{1}{(2\pi)^2 R_1R_2}\frac{2\pi n}{T_2}(x^1-y^1)} \;,
\label{eq:Lambda}\eeq
where for later convenience we have included $y^i$, which are position 
coordinates of the D0-brane on the $T^3$ defined up to $2\pi R_i$.

\subsubsection*{Condition on string field}
\indent

Let us make sure that a D0-brane can really exist in the background 
$B_{ij}$.
Because the $B_{ij}$ is defined using the gauge transformation 
(\ref{eq:BCforB}),  the matter field coupling 
to $B_{ij}$ must be defined by the corresponding gauge transformation. 
This matter field is a string field.
In the covering space of $T^3$, that is $R^3$, there are infinite 
mirrors of the D0-brane, which form a lattice. Let us label the D0-branes
as $(a,b,c)$, where $a,b,c$ are
the integers which indicate the position of the D0-branes on the lattice.
There are strings connecting any two D0-branes. 
For strings which starts from $(a,b,c)$ and ends at $(a',b',c')$,
let us introduce a string field operator
\beq
\phi_{(a,b,c)(a',b',c')} \;.
\eeq
This string field must satisfy the boundary condition corresponding to 
that of $B_{ij}$.

In general, the gauge transformation of string field or string wave 
functional can be 
decided by requiring the invariance of string transition amplitudes.
Let us consider a open string whose end points are on general D-branes.
The transition amplitude is written by
\beq
 \int [{\cal D}X {\cal D}\psi] \;\Psi^{\ast}(c_2,t_2) \Psi(c_1,t_1)\;
e^{i S_{\rm string}} \;,
\eeq
where $\Phi(c_i,t_i)$ are  string wave functionals for string paths 
$c_i$. When $B$ transforms as
\beq
B \to B+\Lambda'^{(1)} \;,
\eeq
the string wave functionals and 1-form gauge fields $A$ on the D-branes 
have to transform as
\beqy
\Psi(c)&\to& \exp(-i T_2 \int_c \Lambda'^{(1)})\; \Psi(c) \\
A &\to& A-\Lambda'^{(1)} 
\eeqy
to keep the amplitude invariant.

In the case of the background we are considering,
corresponding to eq.(\ref{eq:BCforB}), the string field has to satisfy 
the boundary condition
\beq
\phi_{(a,b,c+1)(a',b',c'+1)}=
\exp(-i T_2 \int_{\rm path} \Lambda^{(1)} )\;
\phi_{(a,b,c)(a',b',c')} \;.
\label{eq:BCforSF1}\eeq
Since we are considering the low energy limit, the path can be taken as 
the straight path connecting the two D0-branes.
Substituting eq.(\ref{eq:Lambda}) into eq.(\ref{eq:BCforSF1}), we obtain
\beq
\phi_{(a,b,c+1)(a',b',c'+1)}=e^{i\pi n(a+a')(b-b')}\; 
\phi_{(a,b,c)(a',b',c')}
\; .
\label{eq:BC_condition_c}\eeq
For the translation in $x^1$ or $x^2$ direction, we don't need any gauge
transformation, then
\beq
\begin{array}{c}
\phi_{(a+1,b,c)(a'+1,b',c')}= \phi_{(a,b,c)(a',b',c')} \\
\phi_{(a,b+1,c)(a',b'+1,c')}=\phi_{(a,b,c)(a',b',c')}
\end{array}
\label{eq:BC_condition_b}
\eeq
must be satisfied.
We can easily see that these conditions (\ref{eq:BC_condition_c})
and (\ref{eq:BC_condition_b}) are consistent each other.
Therefore we can say that the background under consideration is really 
consistent.
Note here that in order for the consistency to hold,
$n$ has to be quantized as a integer.

\section{Constraints for the compactification}
\label{sec:constraints}
\label{qc}
\indent

In this section we will consider how to realize the the system 
specified in the last section in Matrix theory.
 We assume that this system can be described by 
Matrix theory with some constraints similar to eq.(\ref{eq:usual_QC}).
 We will examin what constraint is necessary from the correspondence to 
 string theory.

Naively off diagonal matrix elements of Matrix theory correspond to the
components of the string field as
\beq
X^{I}_{(a,b,c)(a',b',c')} \stackrel{?}{\longleftrightarrow}
 \phi^{I}_{(a,b,c)(a',b',c')} \;,
\eeq
but they are not exactly equivalent when background $B$ field exists.
In \cite{CK}, in the case of constant $B$ field on $T^2$, 
the authors showed that
 ordinary products between string fields must be replaced with the 
 $*$-products.
Essentially, the $*$-product is equivalent to  the noncommutative relation 
(\ref{eq:NC_algebra}).On the other hand, in the case of non-constant 
$B$ field, the same argument cannot be applied, but we will see that we
can consider in a similar way.

\subsubsection*{Interaction term}
\indent

In the Matrix theory action (\ref{eq:MT_action}), there are products 
of four off diagonal matrix elements. 
Then, let us consider the product
\beq
I_4=X^{I_1}_{(a_1,b_1,c_1)(a_2,b_2,c_2)} 
X^{I_2}_{(a_2,b_2,c_2)(a_3,b_3,c_3)}
 \cdots X^{I_4}_{(a_4,b_4,c_4)(a_1,b_1,c_1)} \;.
\eeq
The corresponding  four open strings form a closed path,
 and we interpret that this term 
describes an interaction between the four open strings 
(figure \ref{fig:4strings}).
Then this term should correspond to
\beq
I'_4=f\; \phi^{I_1}_{(a_1,b_1,c_1)(a_2,b_2,c_2)}
\phi^{I_2}_{(a_2,b_2,c_2)(a_3,b_3,c_3)} \cdots
\phi^{I_4}_{(a_4,b_4,c_4)(a_1,b_1,c_1)} \;,
\label{eq:I'4}\eeq
where we have included an unknown additional factor $f$ to represent 
the correction by the $B$ field. We will discuss this correction in 
the following.

\setlength{\unitlength}{0.5mm}
\begin{figure}[h]
\begin{center}
\begin{picture}(100,80)(-20,10)
 \put(20,20){\circle*{3}}
 \put(20,20){\line(3,2){30}}
 \put(20,20){\vector(3,2){15}}
 \put(50,40){\circle*{3}}
 \put(50,40){\line(1,4){10}}
 \put(50,40){\vector(1,4){5}}
 \put(60,80){\circle*{3}}
 \put(60,80){\line(-1,0){60}}
 \put(60,80){\vector(-1,0){30}}
 \put(0,80){\circle*{3}}
 \put(0,80){\line(1,-3){20}}
 \put(0,80){\vector(1,-3){10}}
 \put(5,10){$(a_1,b_1,c_1)$}
 \put(55,37){$(a_2,b_2,c_2)$}
 \put(65,78){$(a_3,b_3,c_3)$}
 \put(-38,78){$(a_4,b_4,c_4)$}
\end{picture}
\end{center}
\caption{Interaction of four open strings}
\label{fig:4strings}
\end{figure}
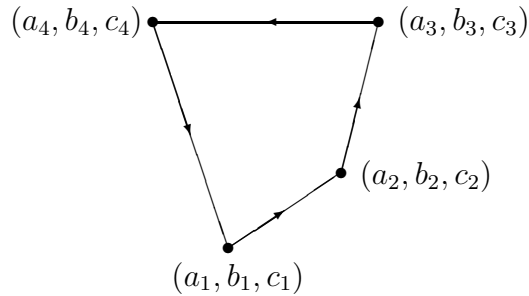
Let us consider a string world sheet $\Sigma$ whose boundary is given by 
the four open strings.
When we move the D0-brane on $T^3$ a little in $x^3$ direction, 
that is $y^3 \to y^3+\Delta y^3$, the world sheet shifts as
$\Sigma \to \Sigma_{\rm new}$ keeping its shape.
The value of the sting action evaluated on the world sheet changes as
\beqy
\Delta S_{\rm string}&=&-T_2 \left[\int_{\Sigma_{\rm new}}B 
-\int_{\Sigma}B \right] \nonumber \\
    &=&- \int_{\Sigma} dx^1 dx^2 
    \frac{n}{(2\pi)^2 R_1R_2R_3} \Delta y^3 \;,
\label{eq:delta_S_}
\eeqy
where we have substituted the explicit form of $B$.
Here the important point is that  $\Delta S_{\rm string}$ is decided 
only by the boundary of $\Sigma$ because the integrand 
in eq.(\ref{eq:delta_S_}) is  constant.
 Since $S_{\rm string}$ changes by $\Delta S_{\rm string}$,
  $I'_4$ should have  dependence on $y^3$ as
\beq
I'_4(y^3+\Delta y^3)=I'_4(y^3)\;e^{i\Delta S_{\rm string}} \;,
\label{eq:local_shift_of_I'}\eeq
and the factor $f$ in eq.(\ref{eq:I'4}) should have this $y^3$ dependence.
Eq.(\ref{eq:delta_S_}) is proportional to the area of the tetragon
 projected on $(x^1,x^2)$ plane from the tetragon decided 
by the four open strings, and we have
\beqy
&&\Delta S_{\rm string}=-n \frac{\Delta y^3}{R_3}\frac 12
\large\{ (a_1 b_2- a_2 b_1) +(a_2 b_3- a_3 b_2) 
 \nonumber \\
&& \qquad \qquad
+(a_3 b_4 - a_4 b_3)+ (a_4 b_1 - a_1 b_4) \large\} \;.
\label{eq:delta_S}\eeqy
From eqs.(\ref{eq:local_shift_of_I'}) and (\ref{eq:delta_S}),
 we  see that if $I'_4$ is written as
\beq
I'_4(y^3)\sim
 e^{i n\frac{y^3}{R_3}\frac12 (-a_1 b_2 +a_2 b_1 )}
 \phi^{I_1}_{(a_1,b_1,c_1)(a_2,b_2,c_2)}
  \cdots
 e^{i n\frac{y^3}{R_3}\frac12 (-a_4 b_1 +a_1 b_4)}
 \phi^{I_4}_{(a_4,b_4,c_4)(a_1,b_1,c_1)}
 ,
\label{eq:I'4_expected}\eeq
 it has the desired $y^3$ dependence. Since $I'_4$ should correspond to
$I_4$,  eq.(\ref{eq:I'4_expected})  suggests that
the correspondence between $X^{I}$ and $\phi^{I}$ must be modified
as
\beq
X^{I}_{(a,b,c)(a',b',c')} \leftrightarrow
e^{in\frac{y^3}{R_3} \frac12 (-a b' +a' b)}\phi^{I}_{(a,b,c)(a',b',c')} \;.
\label{eq:correspondence}\eeq
Here we have ambiguity in the additional factor 
$e^{in\frac{y^3}{R_3} \frac12 (-a b' +a' b)}$ in this relation.
For example, if we replace $(-a b' +a' b)$ with $(a+a')(b-b')$
 or $-(a-a')(b+b')$,
$I_4$ still has the same $y^3$ dependence. In these replacements the 
individual matrix elements
$X^{I}_{(a,b,c)(a',b',c')}$ change but the products like
\beq
X^{I_1}(y^3)_{(a_1,b_1,c_1)(a_2,b_2,c_2)} \cdots
X^{I_k}(y^3)_{(a_k,b_k,c_k)(a_1,b_1,c_1)}
\label{eq:product_X}\eeq
don't change at all. Then we consider that these replacements don't occur 
any physical difference.
In addition we can replace $y^3$ with $y^3+$(constant) in 
eq.(\ref{eq:correspondence}).
This shift corresponds to the change $B_{12}\to B_{12}+(\rm constant)$.
From these considerations we adopt the relation (\ref{eq:correspondence}).
In section \ref{sec:connection},
 we will have an argument which supports this relation.
 
Note that the product (\ref{eq:product_X}) produces a desired 
phase factor under
 the global shift of $y^3$, that is  $c_j\to c_j+1$ or  
 $y^3 \to y^3+2\pi R_3 $ in addition to the local shift of $y^3$.
If we shift each $c_j$ as $c_j \to c_j+1$, the product 
produces a factor due to eq.(\ref{eq:BC_condition_c}). 
 This factor is consistent with the change 
of the string action $S_{\rm string}$ under this shift.

The important point of eq.(\ref{eq:correspondence}) is that the center of 
mass coordinate $y^3$ of the D-brane doesn't decouple from the remaining 
degrees of freedom any longer. In original Matrix theory
 this decoupling occurs, but this is not
the case when an $H$ field background exists.
This result is the main nontrivial feature of the compactification with 
an $H$ field background.

\subsubsection*{Constraints}
\indent

Let us summarize the condition on the matrices for the $T^3$ 
compactification under consideration.
$X^{I}$ must depend on $y^3$ as
\beqy
\label{eq:QC1}
&& X^{I}(y^3)_{(a,b,c)(a',b',c')}=
 e^{in\frac{y^3}{R_3}\frac 12 (-a b' +a' b)}
 X^{I}(0)_{(a,b,c)(a',b',c')}\;,\\
&&\qquad{\rm where} \qquad y^3=X^3_{(a,b,c)(a,b,c)} \quad 
({\rm mod}\; 2\pi R_3 )
\nonumber \;.
\eeqy
The off diagonal matrix elements must satisfy  the conditions
\beqy
X^{I}(0)_{(a+1,b,c)(a'+1,b',c')}&=&
 X^{I}(0)_{(a,b,c)(a',b',c')} \nonumber \\
X^{I}(0)_{(a,b+1,c)(a',b'+1,c')}&=&
  X^{I}(0)_{(a,b,c)(a',b',c')}
 \label{eq:QC2} \\
X^{I}(0)_{(a,b,c+1)(a',b',c'+1)}
&=&e^{i\pi n (a+a')(b-b')} X^{I}(0)_{(a,b,c)(a',b',c')}  \nonumber \;.
\eeqy
For the diagonal matrix elements, we have the conditions
\beqy
X^{I}(0)_{(a+1,b,c)(a+1,b,c)}&=&
 X^{I}(0)_{(a,b,c)(a,b,c)}+2\pi R_1 \delta_{I,1} \nonumber \\
X^{I}(0)_{(a,b+1,c)(a,b+1,c)}&=&
 X^{I}(0)_{(a,b,c)(a,b,c)}+2\pi R_2 \delta_{I,2}
 \label{eq:QC3} \\
X^{I}(0)_{(a,b,c+1)(a,b,c+1)}&=&
 X^{I}(0)_{(a,b,c)(a,b,c)}+2\pi R_3 \delta_{I,3} \nonumber \;.
\eeqy
Note here that the relation
\beq
X^{I}(y^3+2\pi R_3)_{(a,b,c)(a',b',c')}=
 X^{I}(y^3)_{(a,b,c+1)(a',b',c'+1)}
\eeq
is satisfied.
As for the fermionic part $\Theta_{(a,b,c)(a',b',c')}$,
we have  conditions similar to eqs.(\ref{eq:QC1}) and (\ref{eq:QC2}) 
for both off-diagonal and diagonal matrix elements.

\subsubsection*{Solutions to the constraints}
\indent

Let us consider the general solution to the conditions (\ref{eq:QC1})
 $\sim$ (\ref{eq:QC2}).
Using the conditions, any off diagonal elements can be expressed as
\beq
X^{I}(y^3)_{(a,b,c)(a',b',c')}=
 e^{i n \pi (a-a')(b-b') c'} 
 e^{i n \frac{y^3}{R_3} \frac 12 \left[ -(a-a')b' +(b-b')a' \right] }\;
X^{I}(0)_{(a-a',b-b',c-c')(0,0,0)} \;.
\eeq
Here, we used $e^{i n \pi (a-a')(b-b') c'}$ instead of 
$e^{i n \pi (a+a')(b-b') c'}$.
Let us represent 
$X^{I}$  as  operators on a space of functions
\beq
v(\xi_i)=\sum_{a',b',c'} v_{(a',b',c')} e^{ia'\xi_1/\Sigma_1}
e^{ib'\xi_2/\Sigma_2} e^{ic'\xi_3/\Sigma_3} \;,
\eeq
where $v_{(a',b',c')}$ is a vector on which the original matrices 
$X^{I}$ operate, and 
$\Sigma_i=\alpha'/R_i$
are the radii of the ``dual torus"\footnote{We have a space 
dependent $B$ field background, so there is not a T-duality between $T^3
 \leftrightarrow \tilde{T}^3$ from the viewpoint of string theory.}.
The general solution to the conditions (\ref{eq:QC1})
 $\sim$ (\ref{eq:QC2}) can be expressed as
\beq
X^{i}(\xi_j;u^3)=(2\pi\alpha')\left(-i\p^i +A^i(\xi_j;u^3) \right)
,\qquad i=1,2,3 
\label{eq:solution1}\eeq
\beq
A^{i}(\xi_i;u^3)=\sum_{p,q,r} A^i_{(p,q,r)}\;
e^{ip\xi_1/\Sigma_1}e^{iq\xi_2/\Sigma_2} e^{ir\xi_3/\Sigma_3}\;
e^{n \pi pq \Sigma_3 \p_3}\;
e^{n \frac 12 \Sigma_3 u^3 (-p\Sigma_2\p_2 +q\Sigma_1\p_1 ) }
 \; .
\label{eq:solution2}\eeq
Here $u^3$ is related to $y^3$ by
\beq
u^3=y^3/\alpha' =2\pi A^3_{(0,0,0)} \;.
\eeq
As for $X^{a}(\xi_j;u^3)$ $(a=4,\cdots,9)$ and $\Theta(\xi_j;u^3)$, 
they are expressed in the same form as eq.(\ref{eq:solution2}).

\section{Action}
\label{sec:action}
\indent

So far we have considered  the interaction terms in Matrix theory.
Let us now consider the kinetic term 
\beq
{\rm Tr}\bigl[( D_t X^{I})^2 \bigr] \;.
\eeq
Since  $X^{I}$ has  time dependence through $y^3(t)$ as well
as usual one, we have
\beq
\frac{\rm d}{{\rm d}t}X^{I}(t;y^3(t))=
\frac{\p}{\p t} \left.X^{I}(t;y^3(t))\right|_{y^3} + \dot{y^3} 
\frac{\p}{\p y^3} X^{I}(t;y^3(t)) \;.
\label{eq:t-derivative}\eeq
However, due to the second term in eq.(\ref{eq:t-derivative})
and eq.(\ref{eq:QC1}), 
\beq
\frac{\rm d}{{\rm d}t}X^{I}(t;y^3(t))_{(a,b,c)(a',b',c')}
\frac{\rm d}{{\rm d}t}X_{I}(t;y^3(t))_{(a',b',c')(a,b,c)}
\label{eq:product_t}\eeq
turns out to be not invariant under the shift $a,a'\to a+1,a'+1$. 
We have to construct a theory which is invariant under the shifts like 
this because we would like to obtain a theory compactified on $T^3$.
On the other hand if we replace $\frac{\rm d}{{\rm d}t}X^{I}$ with 
$\frac{\p}{\p t}X^{I}|_{y^3}$ in eq.(\ref{eq:product_t}),
the corresponding product turns out to be invariant under these shifts.
This fact suggests that we should interpret $\dot{X}^{I}$ as 
$\frac{\p}{\p t} X^{I}|_{y^3}$.
In the original Matrix theory Lagrangian in flat space,
$\frac{\rm d}{{\rm d} t} X^{I}$
and $\frac{\p}{\p t} X^{I}|_{y^3}$ are equivalent, then we consider that 
there is no contradiction in this interpretation.

 Before we write down the resulting action explicitly,
 let us briefly comment on the generalization to
the case where there are N D0-branes on $T^3$.
In this case we should interpret $y^3$ as the center of mass coordinate 
of the N D0-branes. Each matrix element $X^{I}_{(a,b,c,)(a',b',c')}$ 
is generalized 
to a matrix $X^{I \;kl}_{(a,b,c,)(a',b',c')}$, where $k,l=1,\cdots , N$.
For each $(k,l)$ we have the same conditions as eqs.(\ref{eq:QC1}) 
$\sim$ (\ref{eq:QC3}).

The final Lagrangian can be obtained under the replacements
\beqy
{\rm Tr} &\to& \frac{1}{(2\pi)^3 \Sigma_1 \Sigma_2 \Sigma_3}
\int {\rm d}^3\xi \nonumber \\
X^{I},\; \Theta &\to& X^{I}(\xi,t;u^3),\; \Theta(\xi,t;u^3)\\
\dot{X}^{I},\; \dot{\Theta}
 &\to& {\textstyle \frac{\p}{\p t}} X^{I}(\xi,t;u^3)|_{u^3},\;
 {\textstyle \frac{\p}{\p t}} \Theta(\xi,t;u^3)|_{u^3} \nonumber
\eeqy
in the original Matrix theory Lagrangian (\ref{eq:MT_action}).
We obtain
\beqy
&&S= \frac{1}{g^2_{\rm YM}} \int {\rm d}t{\rm d}^3\xi \;
 {\rm Tr}
 \left\{ -\frac 14 F_{\mu \nu} F^{\mu \nu} 
 -\frac 12 \frac{1}{(2\pi\ap)^2} \bigl( D_{\mu}X^a \bigr)^2
 +\frac 14 \frac{1}{(2\pi\ap)^4} [X^a, X^b]^2 \right. \nonumber \\
 &&\qquad\qquad\qquad \left.
 +\frac 12 \frac{i}{(2\pi\ap)^3} \Theta^T \Gamma^{\mu} D_{\mu}\Theta
 -\frac 12 \frac{1}{(2\pi\ap)^4} \Theta^T \Gamma_a [X^a,\Theta]
 \right\}\;,
\label{eq:final_action}\eeqy
where the Yang-Mills coupling on the dual torus is given by
$
g^2_{\rm YM}=2\pi g_s \Sigma_1 \Sigma_2 \Sigma_3
/\ap^{\frac{3}{2}} \;,
$
and $\mu,\; \nu =0,\cdots, 3$ and
\beqy
 &&F_{\mu \nu}=\p_{\mu} A_{\nu} -\p_{\nu} A_{\mu}
 +i [A_{\mu}, A_{\nu}] \\
 &&D_{\mu}X^a=\p_{\mu} X^a +i [A_{\mu}, X^a] \;, \qquad
 D_{\mu}\Theta=\p_{\mu} \Theta +i [A_{\mu}, \Theta] \;.
\eeqy
The action (\ref{eq:final_action}) appears to be the ordinary  SYM action 
on the dual torus. However, the fields $A_{\mu}$, $X^a$ and $\Theta$
are not ordinary $N\times N$ matrices. Each element of the 
matrices isn't a function but  an operator which is written as
\beq
[A^{i}(\xi_i;u^3)]_{kl}=\sum_{p,q,r} [A^i_{(p,q,r)}]_{kl} \;
e^{ip\xi_1/\Sigma_1}e^{iq\xi_2/\Sigma_2} e^{ir\xi_3/\Sigma_3}\;
e^{n \pi pq \Sigma_3 \p_3}\;
e^{n \frac 12 \Sigma_3 u^3 (-p\Sigma_2\p_2 +q\Sigma_1\p_1 ) }
 \; ,
\eeq
where $[A^i_{(p,q,r)}]_{kl}=[A^i_{(-p,-q,-r)}]^{*}_{lk}$ and
\beq
u^3=2\pi \frac{1}{N}{\rm Tr}[A^3_{(0,0,0)}]
=\frac{1}{(2\pi)^2 \Sigma_1 \Sigma_2 \Sigma_3}
 \int {\rm d}^3\xi \frac{1}{N} {\rm Tr} [A^3(\xi;u^3)] \;.
\eeq
Then, the action (\ref{eq:final_action}) includes a nontrivial 
interaction between the U(1) part of the gauge field $u^3$
and the remaining SU(N) part.

\section{Connection with noncommutative geometry}
\label{sec:connection}
\indent

In this section we will make clear the connection of the constraints 
obtained in section \ref{sec:constraints} with the noncommutative algebra
in the literature\footnote{For example, see \cite{HWW,HW,Ho,MZ}.}
of noncommutative torus compactification.
To do this, we will reformulate the quotient condition into a form
similar to eqs.(\ref{eq:usual_QC}) and (\ref{eq:NC_algebra}).

\begin{figure}[h]
\begin{center}
\begin{picture}(120,100)(-20,-15)
 \put(0,10){\circle*{3}}
 \put(0,10){\line(3,-1){30}}
 \put(0,10.2){\line(3,-1){30}} 
 \put(0,9.8){\line(3,-1){30}} 
 \put(0,10){\vector(3,-1){15}}
 \put(30,0){\circle*{3}}
 \put(30,0){\line(4,1){40}}
 \put(30,0.2){\line(4,1){40}}
 \put(30,-0.2){\line(4,1){40}}
 \put(30,0){\vector(4,1){20}}
 \put(70,10){\circle*{3}}
 \put(70,10){\line(1,2){10}}
 \put(70.2,10){\line(1,2){10}}
 \put(69.8,10){\line(1,2){10}}
 \put(70,10){\vector(1,2){5}}
 \put(80,30){\circle*{3}}
 \put(80,30){\line(-2,3){20}}
 \put(80.2,30){\line(-2,3){20}}
 \put(79.8,30){\line(-2,3){20}} 
 \put(80,30){\vector(-2,3){10}}
 \put(60,60){\circle*{3}}
 \put(60,60){\line(-1,0){30}}
 \put(60,60.2){\line(-1,0){30}}
 \put(60,59.8){\line(-1,0){30}} 
 \put(60,60){\vector(-1,0){15}}%
 \put(30,60){\circle*{3}}
 \put(30,60){\line(-3,-2){30}}
 \put(30,60.2){\line(-3,-2){30}}
 \put(30,59.8){\line(-3,-2){30}} 
 \put(30,60){\vector(-3,-2){15}}
 \put(0,40){\circle*{3}}
 \put(0,40){\line(0,-1){30}}
 \put(0.2,40){\line(0,-1){30}}
 \put(-0.2,40){\line(0,-1){30}} 
 \put(0,40){\vector(0,-1){15}}
 \put(0,10){\line(1,0){70}}
 \put(0,10){\line(4,1){80}}
 \put(0,10){\line(6,5){60}}
 \put(0,10){\line(3,5){30}}
 \put(-30,5){$(a_1,b_1)$}
 \put(25,-13){$(a_2,b_2)$}
 \put(68,-3){$(a_3,b_3)$}
 \put(85,29){$(a_4,b_4)$}
 \put(58,67){$(a_5,b_5)$}
 \put(10,67){$(a_6,b_6)$}
 \put(-30,40){$(a_k,b_k)$}
\end{picture}
\end{center}
\caption{Division of k-polygon into triangles}
\label{fig:k-polygon}
\end{figure}
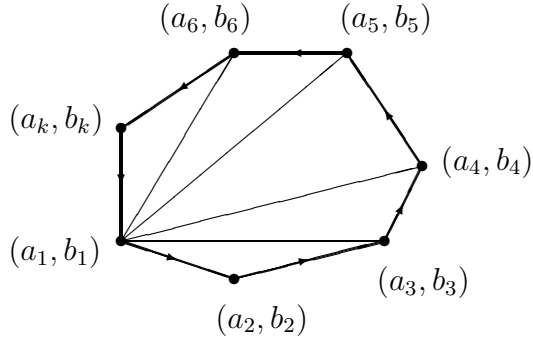

For a while, let us concentrate  our attention only on $(a,b)$
 indices of the matrices
 $X^I(y^3)$, and  consider the trace with respect to $(a,b)$ indices
\beqy
&&\sum_{a_i,b_i} X^{I_1}(y^3)_{(a_1,b_1,c_1)(a_2,b_2,c_2)}
X^{I_2}(y^3)_{(a_2,b_2,c_2)(a_3,b_3,c_3)}\cdots
X^{I_k}(y^3)_{(a_k,b_k,c_k)(a_1,b_1,c_1)} \\ \nonumber
&&\qquad
\equiv \underline{\rm Tr}\left( X^{I_1}(y^3)X^{I_2}(y^3)\cdots
X^{I_k}(y^3) \right) \;.
\label{eq:trace}\eeqy
By bringing together the $y^3$ dependence of all $X^{I_i}(y^3)$, 
the trace can be  written as
\beq
\sum_{a_i,b_i}
X^{I_1}(0)_{(a_1,b_1,c_1)(a_2,b_2,c_2)}\cdots
X^{I_k}(0)_{(a_k,b_k,c_k)(a_1,b_1,c_1)}e^{-i n\frac{y^3}{R_3}{\Delta}}
\eeq
where
\beq
\Delta={\textstyle \frac 12} \left\{(a_1 b_2 -a_2 b_1) 
+(a_2 b_3- a_3 b_2) +\cdots +(a_k b_1-a_1 b_k) \right\} \;,
\eeq
and $\Delta$ is equivalent  to the area of the k-polygon decided by 
the k vertices $(a_i,b_i)$ (figure \ref{fig:k-polygon}).
 Let us change the way to calculate the area and 
divide the k-polygon into triangles
as in figure \ref{fig:k-polygon}. Then $\Delta$ turns into a sum of 
the areas of the triangles,  and can be written as
\beqy
&&\Delta={\textstyle \frac 12}
 \left\{ (a_1-a_2,b_1-b_2)\times(a_2-a_3,b_2-b_3)+
 (a_1-a_3,b_1-b_3)\times(a_3-a_4,b_3-b_4) \right. \nonumber \\
 &&\qquad\qquad \left.+ (a_1-a_4,b_1-b_4)\times(a_4-a_5,b_4-b_5)
 +\cdots \right\} \;,
\eeqy
where $\times$ denotes the operation of the outer product. 

Now let us  introduce $\underline{*}$-product by the relation
\beq
X^{I_1}\underline{\ast}X^{I_2}
\equiv\sum_{a_2,b_2} e^{-i\pi \theta(y^3)
\left[(a_1-a_2,b_1-b_2)\times(a_2-a_3,b_2-b_3)\right]}
X^{I_1}_{(a_1,b_1,c_1)(a_2,b_2,c_2)}
X^{I_2}_{(a_2,b_2,c_2)(a_3,b_3,c_3)} \;,
\eeq
where
\beq
2\pi\theta(y^3)=\frac{ny^3}{R_3}=T_2\int_{T^2(x^3=y^3)}B \;.
\eeq
This $\underline{*}$-product is defined only with respect to $(a,b)$
 indices and we don't carry out the summation with respect to
 $c$ indices, 
 so that we have used the notation $\underline{*}$ instead of usual $*$.
The $\underline{*}$-product is almost same as the ordinary $*$-product.
However, the parameter $\theta(y^3)$ is not just a constant but it is dependent
on the dynamical variable $y^3$.
Using the $\underline{*}$-product, it is not difficult to show that
the trace eq.(\ref{eq:trace}) can be written as
\beq
\underline{\rm Tr}\left( X^{I_1}(y^3)X^{I_2}(y^3)\cdots
X^{I_k}(y^3)\right)
= \sum_{a_1,b_1}  X^{I_1}(0)\underline{\ast}X^{I_2}(0)
\underline{\ast}\cdots \underline{\ast}X^{I_k}(0) \;.
\label{eq:trace_product}\eeq
Note that  all the $y^3$ dependence on the left hand side 
has turned into the $y^3$ dependence in the definition of 
 the $\underline{*}$-product on the right hand side. 
In other words, if we use the $y^3$-dependent $\underline{*}$-products
 instead of the usual products, we can think $X^{I}$ independent of $y^3$.

Since it is known that  the effect of the $\underline{*}$-product 
is equivalent to 
the algebra of $\tilde{U}_i$ satisfying the noncommutative relation
\beq
\begin{array}{rcll}
\tilde{U}_1 \tilde{U}_2 & = & e^{-i 2\pi \theta(y^3)} 
 \tilde{U}_2 \tilde{U}_1 & \\
 \tilde{U}_1 \tilde{U}_3 & = & \tilde{U}_3 \tilde{U}_1\;, &
\tilde{U}_2 \tilde{U}_3 = \tilde{U}_3 \tilde{U}_2 \;,
\label{eq:new_NC_relation}
\end{array}
\eeq
the constraints obtained in section \ref{sec:constraints} can be put into 
a form similar to eq.(\ref{eq:usual_QC}). Let us follow \cite{MZ}, and
rewrite the quotient condition.
Introducing the operators $\p_i$  defined by\footnote{These $\p_i$ don't 
correspond to $\p_i$ in eq.(\ref{eq:solution1}).
 In this section we change the 
normalization to make clear the correspondence, for example, to \cite{MZ}.}
\beq
[\p_i,\tilde{U}_j]=i \tilde{U}_j \delta_{ij}\;,
\eeq
we define the unitary operators $U_i$ by
\beq
U_1=\tilde{U}_1 e^{2\pi \theta (y^3) \p_2}\;,\qquad
U_2=\tilde{U}_2 e^{-2\pi \theta (y^3) \p_1}\;,\qquad
U_3=\tilde{U}_3 \;.
\eeq
Then $U_i$ and $\tilde{U}_j$ commute each other, and $U_i$ satisfy
the noncommutative relation
\beq
\begin{array}{rcll}
 U_1U_2 & = & e^{i2\pi \theta(y^3)} U_2U_1 &   \\
 U_1U_3 & = & U_3U_1\;, &   U_2U_3=U_3U_2 \;.
\end{array}
\label{eq:algebra'}\eeq
Using $U_i$, the quotient condition corresponding to 
eqs.(\ref{eq:QC1}) $\sim$ (\ref{eq:QC3})
can be written as
\beqy
U_1^{-1}X^{I}U_1&=&X^{I}+ 2\pi R_1\delta_{I,1} \nonumber \\
U_2^{-1}X^{I}U_2&=&X^{I}+ 2\pi R_2\delta_{I,2}
 \label{eq:QC'} \\
U_3^{-1}X^{I}U_3&=&\left[e^{-i\pi n \overline{\p}_1\overline{\p}_2}
X^{I} \right]+ 2\pi R_3\delta_{I,3}\;. \nonumber
\eeqy
Here,
$
y^3=\frac{1}{N}{\rm Tr}\left( X^3 \right)
\; ( {\rm mod}\;\; 2\pi R_3)
$,
and  we introduced the operators $\overline{\p}_i$  which are  defined by
\beq
\left[ \overline{\p}_i ^m \tilde{U}_1^{p_1} \tilde{U}_2^{p_2} 
\tilde{U}_3^{p_3} \right]
=(ip_i) ^m \tilde{U}_1^{p_1} \tilde{U}_2^{p_2} \tilde{U}_3^{p_3} \;.
\eeq
In the third equation of (\ref{eq:QC'}) we have introduced the
operation $[e^{-i\pi n \overline{\p}_1\overline{\p}_2}
\cdots ]$,  so as to incorpolate the effect of the gauge
transformation of the string field, eq.(\ref{eq:BC_condition_c}).
So  the quotient condition eq.(\ref{eq:QC'}) is a little different from
that of  $T^d$ compactification with constant B-field, 
eq.(\ref{eq:usual_QC}).

 In units where $2\pi\ap =1$, the general solution to eq.(\ref{eq:QC'}) 
 corresponding to the trivial bundle 
is given by
\beqy
&&X^i=-2\pi i R_i \p_i +[ e^{-\pi n \overline{\p}_1 \overline{\p}_2 \p_3}
 A^i(\tilde{U}_j) ]\;, \qquad i=1,2,3 
\label{eq:another_solution} \\
&&X^a=[ e^{-\pi n \overline{\p}_1 \overline{\p}_2 \p_3}
 X'^a(\tilde{U}_j) ]\;, \qquad a=4,\cdots,9 \;,
\eeqy
where  $A^i(\tilde{U}_j)$ and  $X'^a(\tilde{U}_j)$ are arbitrary power 
functions of $\tilde{U}_j$.
In order to make clear the correspondence of the description
in this section to that in section 3,
let us compare eq.(\ref{eq:another_solution}) with eqs.(\ref{eq:solution1}) 
and (\ref{eq:solution2}).
The effect of the factor 
$e^{n\frac12 \Sigma_3 u^3 (-p \Sigma_2\p_2+ q\Sigma_1\p_1) }$
in eq.(\ref{eq:solution2}) is realized through the $y^3$-dependent
 noncommutative relation (\ref{eq:new_NC_relation}).
But the factor 
$e^{n\pi pq \Sigma_3 \p_3}$ in eq.(\ref{eq:solution2}) still remains and
it has turned into the operation 
$[ e^{-\pi n \overline{\p}_1 \overline{\p}_2 \p_3}\cdots ]$
in eq.(\ref{eq:another_solution}).

The quotient condition (\ref{eq:QC'}) can also be  applied to the case of
 N D0-branes. Although we don't discuss solutions corresponding to 
 twisted bundles explicitly, we expect that they also exist as in
 \cite{Ho,MZ}.

\section{Conclusion}
\label{sec:conclusion}
\indent

In this paper, we considered D0-branes compactified on $T^3$ with
 an H-field background.
We assumed that this system can be described by Matrix theory
with  some appropriate constraints on the matrices.
 Then we examined what constraints we 
should impose in order for the resulting theory to be consistent with
string theory. We obtained eqs.(\ref{eq:QC1}) $\sim$ (\ref{eq:QC3})
as the constraints. 
We put the constraints into a form where the noncommutative relation is
apparent, eqs.(\ref{eq:algebra'}) and (\ref{eq:QC'}).

The resulting theory is not ordinary super Yang-Mills theory
on a noncommutative three-torus as in the case
of constant background B-field.
In the case of an H-field background, the parameter of 
 noncommutativity $\theta$ is not a constant, but it 
becomes $y^3$-dependent and dynamical.
Thus, in general, the center of mass 
coordinates of the D0-branes don't decouple from the remaining degrees
of freedom any longer.
We also have to introduce an additonal operation
$[e^{-\pi n \overline{\p}_1 \overline{\p}_2 \p_3} \cdots ]$
in the gauge field on the dual torus, eq.(\ref{eq:another_solution}),
in order to incorpolate the effect of the gauge transformation
of the string field, eq.(\ref{eq:BC_condition_c}).
 These are the main differences from the
 compactification with constant B-field.

We investigated this system as a first step to examine the possibility
to describe transverse 
M5-branes  in Matrix theory. Transverse M5-branes correspond to NS5-branes 
in type IIA sting theory.
 Does this system have a corresponding 5-brane charge?
Let us consider the relation of the background we discussed and NS-5branes.
For example, for $n=1$ case the background corresponds to the configuration
where oppositely charged two NS5-branes are infinitely 
away from each other in one of the directions transverse to $T^3$.
Thus the total brane charge is zero, and we can't expect nonzero
brane charge.
We consider, however, that it will be very interesting to examine whether
the discussion in this paper can be extended to the case of a background
with nonzero 5-brane charge.

It would also be interesting to examine the relation of
the present formalism to the matrix theory
in weak background fields \cite{TR} and supermembrane theory in 
an arbitrary SUGRA background \cite{DWPP}.

\section*{Acknowledgements}
\indent

We would like to thank H. Kunitomo for helpful discussions and careful
reading of the manuscript.
We are grateful Prof. M. Ninomiya for warmhearted encouragement.
This work was supported in part by
JSPS Research Fellowship for Young Scientists and
 the Grant-in-Aid for Scientific
Research (8970) from the Ministry of Education, Science and Culture.

\end{document}